\documentclass[prb,floatfix,superscriptaddress,twocolumn,amsmath,amssymb]{revtex4}

\usepackage{graphicx}
\usepackage{dcolumn}
\usepackage{bm}
\usepackage{amssymb}
\usepackage{amsmath}
\usepackage{color}
\usepackage{hyperref}

\newcommand{\pr}{^{\prime}}

\begin{document}

\bibliographystyle{apsrev}

\title{Entanglement entropy and multifractality at localization transitions}

\author{Xun Jia}
\affiliation{Department of Physics and Astronomy, University of California Los Angeles, Los Angeles, California,
90095-1547}

\author{Arvind R.~Subramaniam}
\affiliation{James Franck Institute and Department of Physics, University of Chicago, Chicago, Illinois, 60637}

\author{Ilya A.~Gruzberg}
\affiliation{James Franck Institute and Department of Physics, University of Chicago, Chicago, Illinois, 60637}

\author{Sudip Chakravarty}
\affiliation{Department of Physics and Astronomy, University of California Los Angeles, Los Angeles, California,
90095-1547}

\date{December 20, 2007}

\begin{abstract}
The von Neumann entanglement entropy is a useful measure to characterize a quantum phase transition. We investigate the
nonanalyticity of this entropy at disorder-dominated quantum phase transitions in noninteracting electronic systems. At
these critical points, the von Neumann entropy is determined by the single particle wave function intensity, which
exhibits complex scale invariant fluctuations. We find that the concept of multifractality is naturally suited for
studying von Neumann entropy of the critical wave functions. Our numerical simulations of the three dimensional
Anderson localization transition and the integer quantum Hall plateau transition show that the entanglement at these
transitions is well described using multifractal analysis.
\end{abstract}


\maketitle

\section{Introduction}
Entanglement is a unique feature of a quantum system and entanglement entropy, defined through the von Neumann entropy
(vNE) measure, is one of the most widely used quantitative measures of entanglement.
\cite{Osborne2002,Calabrese2004,Vidal2004,Kopp2007} Consider a composite system that can be partitioned into two
subsystems $A$ and $B$. The vNE of either of the subsystems is $s_{A}=-\mathrm{Tr}_{A}\rho_{A}\ln \rho_{A}=s_{B}=
-\mathrm{Tr}_{B}\rho_{B}\ln \rho_{B}$. Here, the reduced density matrix $\rho_{A}$  is  obtained by tracing over the
degrees of freedom in $B$: $\rho_{A}=\mathrm{Tr}_{B}|\psi_{A B}\rangle\langle\psi_{AB}|$ and similarly for $\rho_{B}$.
In general, for a pure state $|\psi_{A B}\rangle$ of a composite system, the reduced density matrix is a mixture, and
the corresponding entropy is a good measure of entanglement.

The scaling behavior of entanglement entropy is a particularly useful characterization near a quantum phase transition
\cite{Vidal2004}. The entanglement entropy can show nonanalyticity at the phase transition even when the ground state
energy (the quantum analog of the classical free energy) is analytic. While these ideas have been studied in a number
of translation-invariant models,\cite{Vidal2004,Wu2004,Calabrese2004} there have been far fewer investigations of
random quantum critical points (for notable exceptions, see Ref.~\onlinecite{Refael2004}).

In particular, noninteracting electrons moving in a disordered potential can undergo continuous quantum phase
transitions between an extended metallic and a localized insulating state as the Fermi energy is varied across a
critical energy $E_C$. Well known examples are the Anderson transition in three dimensions and the integer quantum Hall
(IQH) plateau transition in two dimensions where the ground state energy does not exhibit any nonanalyticity. In
contrast, vNE will be shown to exhibit nonanalyticity at these transitions and a scaling behavior. At the outset, it
should be emphasized that because of the single particle and disorder-dominated nature of these quantum phase
transitions, entanglement as characterized by vNE and its critical scaling behavior are fundamentally different from
those calculated for interacting systems. This statement will be made more precise later.

In a noninteracting electronic system close to a disordered critical point, the wave function intensity at energy $E$,
$|\psi_E(r)|^2$, fluctuates strongly at each spatial point $r$ and, consequently, has a broad (non-Gaussian)
distribution even in the thermodynamic limit.\cite{Castellani1986} This non-self-averaging nature of the wave function
intensity is characterized through the scaling of its moments. In particular, moments of normalized wave function
intensity, $P_q$ (called the generalized inverse participation ratios), obey the finite-size scaling ansatz,
\begin{align} \label{definition of Pq}
P_q(E) \equiv \sum_{r} \overline{ \left| \psi_E(r)\right|^{2q}}
\sim L^{-\tau_q} \, \mathcal{F}_q\big[(E-E_C)L^{1/
\nu}\big].
\end{align}
Here, $L$ is the system size, $\nu$ is the exponent characterizing the divergence of correlation length, $\xi_E \sim
|E-E_C|^{-\nu}$. $\tau_q$ is called the multifractal spectrum, and the overbar denotes averaging over different
disorder realizations. $\mathcal{F}_q(x)$ is a scaling function with $\mathcal{F}_q(x\rightarrow 0) = 1$ close to the
critical point $E=E_C$. When $E$ is tuned away from $E_C$, the system either tends towards an ideal metallic state with
$P_q(E) \sim L^{-D(q-1)}$ ($D$ being the number of spatial dimensions) or becomes localized with $P_q(E)$ independent
of $L$.

Below, we first show that the disorder-averaged vNE can be expressed as a derivative of $P_q$ and thus, its scaling
behavior follows from multifractal analysis. After that, we apply our formalism to understand the numerical results on
vNE at the three dimensional Anderson localization and IQH plateau transitions. vNE in the Anderson localization
problem was studied previously,\cite{Kopp2007,Varga2007} but the connection with mulitfractality and the unique
features of vNE at these quantum phase transitions have not been clearly elucidated.

\section{Entanglement Entropy in Disordered Noninteracting Electronic Systems}

Even though the disorder induced localization problem can be studied in a single particle quantum mechanics language,
there is no obvious way to define entanglement entropy in this picture. However (see Ref.~\onlinecite{Zanardi2002}),
entanglement can be defined using the site occupation number basis in the second-quantized Fock space. Let us divide
the lattice of linear size $L$ into two regions, $A$ and $B$. A single particle eigenstate of a lattice Hamiltonian at
energy $E$ is represented in the site occupation number basis as
\begin{align}
| \psi_E \rangle &= \sum _{r \in A \cup B} \psi_E(r) \, |1
\rangle_r \bigotimes_{r\pr \ne r } \, |0 \rangle_{r\pr}
\end{align}
Here $\psi_E(r)$ is the normalized single particle wave
function at site $r$ and $|n\rangle_r$ denotes a state having
$n$ particles at site $r$. We decompose the above sum over
lattice sites $r$ into the mutually orthogonal terms,
\begin{align}\label{decomposition of state}
| \psi_E \rangle = |1 \rangle_A \otimes |0 \rangle_{B} + |0
\rangle_A \otimes |1 \rangle_{B}
\end{align}
where
\begin{align}
|1 \rangle_A &= \sum _{r \in A} \psi_E(r) |1 \rangle_r
\bigotimes_{r\pr  \ne r } |0 \rangle_{r\pr}, \,|0 \rangle_A =
\bigotimes_{r \in A } |0 \rangle_{r}
\end{align}
with analogous expressions for the $|1 \rangle_B$ and $|0
\rangle_B$ states. Notice that these states have the
normalization
\begin{align}
\langle 0|0 \rangle_A = \langle 0|0 \rangle_B = 1, \, \langle
1|1 \rangle_A = p_A, \, \langle 1|1 \rangle_B = p_B,
\end{align}
where
\begin{align}
 p_{A}= \sum_{r \in A}  |\psi_E(r)|^2,
\end{align}
and similarly for $p_B$ with $p_A + p_B =1$.

To obtain the reduced density matrix $\rho_A$, we trace out the
Hilbert space over $B$ in the density matrix $\rho = | \psi_E
\rangle \langle \psi_E |$. This gives,
\begin{align}
\rho_A  & = |1 \rangle_A \langle 1| + p_B |0 \rangle_A \langle 0|.
\end{align}
The corresponding vNE is given by
\begin{align}\label{bipartite entanglement}
s_A = - p_A \ln p_A - p_B \ln p_B.
\end{align}
In the above equation, we see that manifestly $s_A = s_B$. More importantly, $s_A$ is bounded between $0$ and $\ln 2$
for any eigenstate. This is in sharp contrast to the entanglement entropy in interacting quantum systems where it can
be arbitrarily large near the critical point. The reason for this is also clear: Even though we used a second-quantized
language, we are dealing with a single particle state rather than a many body correlated state. Consequently, the
entanglement entropy does not grow arbitrarily large as a function of the size of $A$.

We also observe that at criticality, if the whole system size becomes very large in comparison with the subsystem $A$,
we can restrict the subsystem to be a single lattice site and study the scaling dependence with respect to the overall
system size $L$. Then, using the ansatz of scale invariance, we can always find the scaling of the entanglement as a
function of the subsystem size $l$ since near criticality, only the dimensionless ratio $L/l$ can enter any physical
quantity. To extract scaling, we find the bipartite entanglement of a single site $r$ with the rest of the system and
sum this over all lattice sites in the system. Using Eq. \eqref{bipartite entanglement}, we write this as
\begin{align}\label{single site entropy sum}
S(E)  &= - \sum_{r \in L^d} \Bigl\{ |\psi_E(r)|^2   \ln |\psi_E(r)|^2 \nonumber \\ & \quad + \left[1- |\psi_E(r)|^2
\right] \ln \left[ 1- |\psi_E(r)|^2 \right]\Bigr\}.
\end{align}

To leading order, the second term inside the square bracket in Eq. \eqref{single site entropy sum} can be dropped since
$\left| \psi_{E}(r)\right|^2  \ll 1$ at all points $r$ when the states are close to the critical energy. We can readily
relate the disorder average (denoted by overbar) of this entropy to the multifractal scaling in Eq. \eqref{definition
of Pq} and get the $L$ scaling as
\begin{align}\label{EE summed over all sites}
\overline{S}(E)\approx -\frac{dP_{q}}{dq}\bigg|_{q=1} \approx
\frac{d\tau_q}{dq}\bigg|_{q=1} \ln L - \frac{\partial
\mathcal{F}_q}{\partial q} \bigg|_{q=1}.
\end{align}
We do not know the general form of the scaling function $\mathcal{F}_q$, but we can get the approximate $L$ dependence
of the entropy in various limiting cases. For the exactly critical case when $\mathcal{F}_q \equiv 1$ for all values of
$q$, we get
\begin{align}\label{EE critical scaling}
\overline{S}(E) \sim \alpha_1 \ln L,
\end{align}
where  the constant $\alpha_1 ={d\tau_{q}/dq}|_{q=1}$
%at criticality and
is unique for each universality class. From the discussion following Eq. \eqref{definition of Pq}, the leading scaling
behavior of $\overline{S}(E)$ in the ideal metallic and localized states is given by $D \ln L$ and $\alpha_1 \ln
\xi_E$, respectively. From the limiting cases, we see that, in general, $\overline{S}(E)$ has the approximate form
\begin{align}\label{approximate form for EE of single energy state}
\overline{S}(E) \sim \mathcal{K}[(E-E_C)L^{1/\nu}] \ln L,
\end{align}
where the coefficient function $\mathcal{K}(x)$  decreases from
$D$ in the metallic state to $\alpha_1$ at criticality and then
drops to zero for the localized state. We will see that this
scaling form is verified in our numerical simulations.

\section{Entanglement in the three dimensional Anderson Model}

The scaling form for the entanglement entropy averaged over all eigenstates of the single particle Hamiltonian is also
of interest since this scaling can change as a function of disorder strength. To be specific, let us consider the 3D
Anderson model on a cubic lattice. The Hamiltonian is
\begin{equation}\label{Hamiltonian_Anderson}
    H=\sum_i V_i c_i^\dag c_i-t\sum_{\langle i,j\rangle}(c_i^\dag
    c_j+H.c.),
\end{equation}
where $c_i^{\dag}$($c_i$) is the fermionic creation (annihilation) operator at site $i$ of the lattice, and the second
sum is over all nearest neighbors. We set $t=1$, and the $V_i$ are random variables uniformly distributed in the range
$[-W/2,W/2]$.  It is known \cite{MacKinnon1981} that as $W$ is decreased from a very high value, extended states appear
at the band center below the critical disorder strength $W_c=16.3$, and a recent work \cite{Slevin2001} reported the
localization length exponent $\nu=1.57\pm 0.03$.

The analysis leading to Eq. \eqref{approximate form for EE of
single energy state} also holds when we study wave functions at
a single energy, say $E=0$ and increase the disorder strength
in the Anderson model across the critical value $W_c$. In this
case, the states at $E=0$ evolve continuously from fully
metallic to critical and then finally localized, resulting in
the approximate form for the entanglement entropy as
\begin{align}\label{generalscaling2}
\overline{S}(E=0,w,L) \sim \mathcal{C}(wL^{1/\nu})\ln L,
\end{align}
where  $w=(W-W_c)/W_c$ is the normalized relative disorder
strength and $\mathcal{C}(x)$ is a scaling function. In
particular, as mentioned before, $\mathcal{C}(x) \to D$ as $w
\to -1$, $\mathcal{C}(x) \to 0$ as $w \to \infty$, and
$\mathcal{C}(x) =\alpha_1$ when $w=0$.

Next, we look at the energy-averaged entropy. We average Eq.
\eqref{EE summed over all sites} over the  entire  band of
energy eigenvalues and construct the vNE,
\begin{align}
\overline{S}(w,L) = \frac{1}{L^3}\sum_{E} \overline{S}(E,w,L),
\end{align}
where $L^3$ is also the total number of states in the band. Then using Eqs. \eqref{approximate form for EE of single
energy state} and \eqref{generalscaling2}, one can show that close to $w=0$,
\begin{align}\label{generalscaling1}
\overline{S}(w,L) \sim C +L^{-1/\nu}f_{\pm}\big(wL^{1/\nu}\big)\ln L,
\end{align}
where $C$ is an \emph{L independent} constant, and $f_\pm(x)$
are two universal functions corresponding to the two regimes
$w>0$ and $w<0$.

\begin{figure}
    \centering
    \includegraphics[width=\columnwidth]{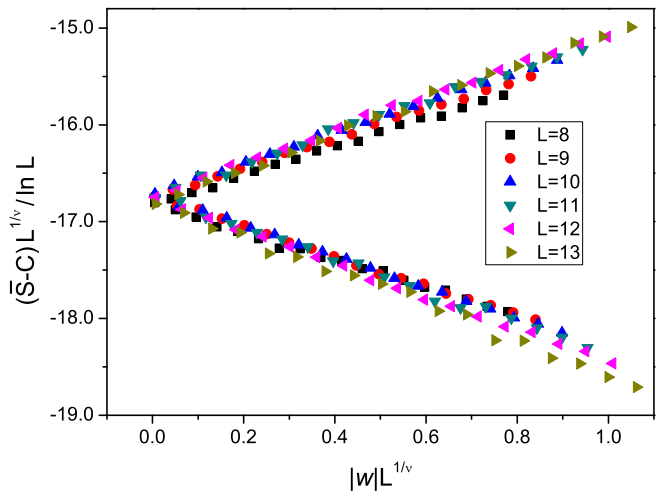}\\
    \caption{(Color online) Scaling curve in the 3D Anderson model. With the
    choice of $\nu=1.57$ and $C=12.96$, all data collapse to a
    universal functions $f_\pm(x)$. The two branches correspond
    to $w<0$ and $w>0$.} \label{Scaling_Anderson1}
\end{figure}

We numerically diagonalize the Hamiltonian [Eq. \ref{Hamiltonian_Anderson}] in a finite $L\times L\times L$ system with
periodic boundary conditions.  The maximum system size is $L=13$, and the results are averaged over 20 disorder
realizations. The scaling form of $\overline{S}(w,L)$ is given by Eq. \eqref{generalscaling1}.
Figure~\ref{Scaling_Anderson1} shows the results of the data collapse with a choice of $\nu=1.57$, and the nonuniversal
constant $C=12.96$ is determined by a powerful algorithm described in Ref.~\onlinecite{Goswami2007}. The successful
data collapse reflects the nonanalyticity of the von Neumann entropy and accuracy of the multifractal analysis.

We also use the transfer matrix method \cite{Kramer1996} to study  the energy dependence of $\overline{S}(E,w,L)$ by
considering a quasi-one-dimensional (quasi-1D) system with a size of $(mL)\times L\times L$, $m\gg 1$.   We use $L$ up
to $18$, and $m=2000 \gg 1$ is found to be sufficient. To compute vNE, we divide the quasi-1D system into $m$ cubes
labeled by $I=1,2,\ldots,m$, each containing $L^3$ sites. We normalize the wave function within each cube and compute
the vNE, $\overline{S^I}(E,W,L)$, in the $I^{\text{th}}$ cube, and finally $\overline{S}(E,W,L)$ is obtained by
averaging over all cubes.

\begin{figure}
\centering
  \includegraphics[width=\columnwidth]{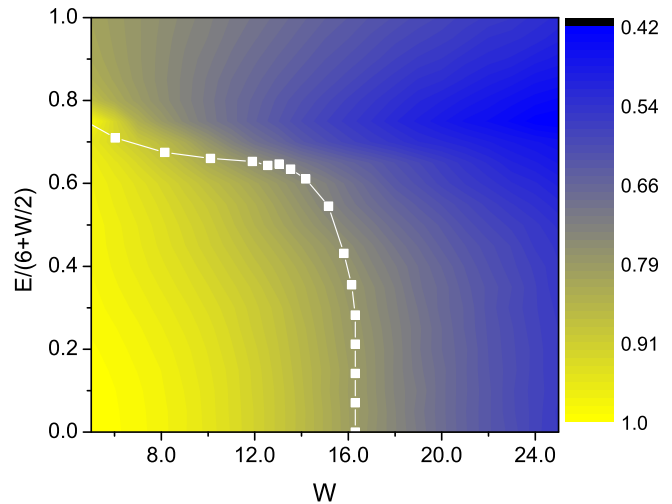}\\
  \caption{(Color online) $\overline{S}(E,W,L)$ as a function of $E$ and $W$
  computed in a system with $L=10$. The square shows the
  mobility edge reported in Ref.~\onlinecite{Bulka1987}.
  Because of the finiteness of the system, the transition
from the localized to the delocalized region is smooth.}
  \label{S_EW}
\end{figure}

\begin{figure}
\centering
  \includegraphics[width=\columnwidth]{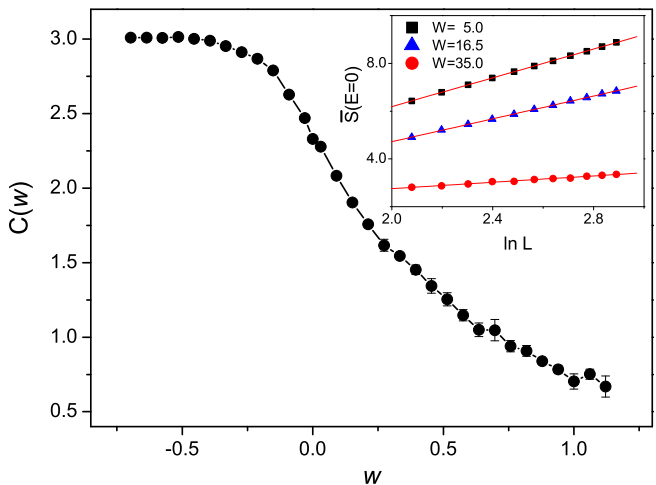}\\
  \caption{(Color online) The quantity ${\cal C}$ in
  Eq.~(\ref{generalscaling2}). The range of  the system sizes
  is too small to observe the weak $L$ dependence. Inset:
$\overline{S}(E=0,W,L)$ as a function of $\ln L$ for three different $W$.}
  \label{scaling_Anderson2}
\end{figure}

A typical  $\overline{S}(E,W,L)$  with $L=10$ is shown in Fig.~\ref{S_EW}. The value of $\overline{S}(E,W,L)$ is
normalized by $\ln(L^3)$ such that $\overline{S} \to 1$ in a fully extended state. The energy $E$ is normalized by
$(W/2+6)$, which is the energy range of nonzero density of states.\cite{Wegner1981}  The mobility edge computed in
Ref.~\onlinecite{Bulka1987} is also plotted in Fig.~\ref{S_EW}. The validity of the scaling form in
Eq.~(\ref{generalscaling2}) is seen in Fig.~\ref{scaling_Anderson2}. In particular, the function $\mathcal C(x)$ shows
the expected behavior.

%The inset shows $\overline{S}(E=0,W,L)$ as a function of $\ln
%L$ for $3$ different $W$ .

\section{Entanglement in the integer quantum Hall system}

Consider now the second example, the integer quantum Hall system in a magnetic field $B$. The Hamiltonian can be
expressed \cite{Huckestein1995} in terms of the matrix elements of the states $|n,k\rangle$, where  $n$ is the  Landau
level index and  $k$ is the wave vector in the $y$ direction. Focussing on the lowest Landau level $n=0$, with the
impurity distribution $\overline{V(\mathbf{r})V(\mathbf{r'})}=V_0^2\delta(\mathbf{r}-\mathbf{r'})$, the matrix element
$\langle 0,k|V|0,k'\rangle$ can be generated as in Ref.~\onlinecite{Huckestein1995}.

Now, consider a two dimensional square with a linear dimension $L=\sqrt{2\pi} Ml_B$, where $l_B=(\hbar/eB)^{1/2}$ is
the magnetic length and $M$ is an integer, with periodic boundary conditions imposed in both directions. We discretize
the system with a mesh of size $\sqrt{\pi} l_B/\sqrt{2}M$. The Hamiltonian matrix  is diagonalized and a set of
eigenstates $\{|\psi_a\rangle=\sum_k \alpha_{k,a}|0,k\rangle\}_{a=1}^{M^2}$ is obtained with corresponding eigenvalues
$\{E_a\}_{a=1}^{M^2}$. The energies are measured relative to the center of the lowest Landau band \cite{Ando1974} in
units of $\Gamma=2V_0/\sqrt{2\pi}l_B$. Finally, for each eigenstate the wave function in real space can be constructed
as
\begin{equation}
    \psi_a(x,y)=\langle x,y|\psi_a\rangle=\sum_k
    \alpha_{k,a}\psi_{0,k}(x,y), \label{wave function_IQHE}
\end{equation}
where $\psi_{0,k}(x,y)$ is the lowest Landau level wave
function with a momentum quantum number $k$.

The dimension of the Hamiltonian matrix increases as $N_k\sim M^2$, making it difficult to diagonalize fully. Instead,
we compute only those states $|\psi_a\rangle$ whose energies lie in a small window $\Delta$ around a preset value $E$,
i.e. $E_a\in[E-\Delta/2,E+\Delta/2]$. We ensure that $\Delta$ is sufficiently small ($0.01$) while at the same time,
there are enough states in the interval $\Delta$ (at least 100 eigenstates).

\begin{figure}
\vskip 5 mm
 \centering
  \includegraphics[width=\columnwidth]{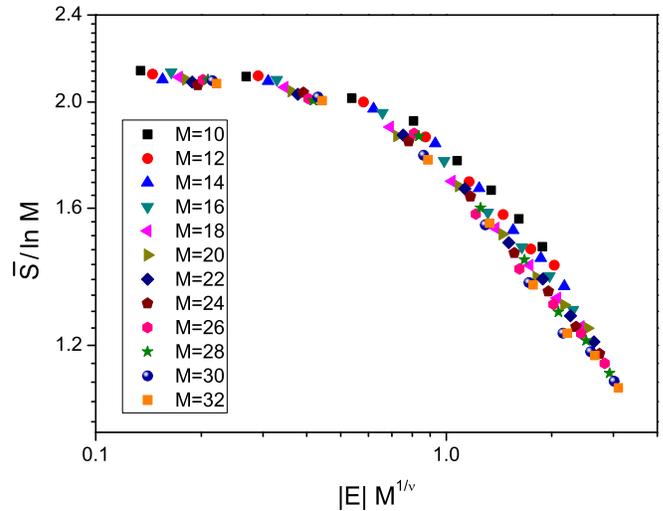}
  \caption{(Color online) Scaling of the von Neumann entropy  $\overline{S}(E)$ for the IQHE. $M$ instead of $L$ is used in the data
  collapse with the accepted value of $\nu=2.33$.}
  \label{scaling_IQHE}
\end{figure}

We now uniformly break up the $L\times L$ square into nonoverlapping squares $\mathcal{A}_i$ of size $l\times l$, where
$l=l_B\sqrt{\pi/2}$, independent of the system size $L$. For each of the states, we compute the coarse grained quantity
$\int_{(x,y)\in\mathcal{A}_i}|\psi_a(x,y)|^2\mathrm{d}x\mathrm{d}y$. The computation of the vNE for a given eigenstate
follows the same procedure described for the Anderson localization. Finally, by averaging over states in the interval
$\Delta$, the vNE $\overline{S}(E, L)$ is obtained at the preset energy $E$. The scaling form of $\overline{S}(E,L)$ is
given by Eq. \eqref{approximate form for EE of single energy state} with $E_C = 0$ and is
$\overline{S}(E,L)=\mathcal{K}(|E|L^{1/\nu})\ln L$. A good agreement with the numerical simulations is seen in
Fig.~\ref{scaling_IQHE}.

\section{Conclusions}

We have clearly established the formalism for computing the entanglement entropy near quantum critical points in
noninteracting disordered electronic systems. We have also identified its relation with the well-studied notion of
multifractality and illustrated our concepts through numerical simulations of two important models, the 3D Anderson
transition and the IQH plateau transition. This work represents a starting point to study entanglement in electronic
systems with both disorder and interactions.

\section{Acknowledgements}

This work was supported by
%the National Science foundation
NSF Grant No. DMR-0705092 (S.C. and X.J.), NSF MRSEC Program under Grant No. DMR-0213745, the NSF Career grant
DMR-0448820 and the Research Corporation (I.A.G. and A.R.S.). A.R.S. and I.A.G. acknowledge hospitality at the
Institute for Pure and Applied Mathematics, UCLA where this work was started. S.C. would also like to thank the Aspen
Center for Physics.

%\bibliography{referencev4}

\end{document}